\documentclass[12pt,twoside]{article}
\usepackage{fleqn,espcrc1}
\usepackage{xspace,amssymb}

\usepackage{graphicx}
\usepackage[figuresright]{rotating}

\newcommand{\Msun}{{\ensuremath{\mathrm{M}_{\odot}}}\xspace}
\newcommand{\erg}{{\ensuremath{\mathrm{erg}}}\xspace}
\newcommand{\K}{{\ensuremath{\mathrm{K}}}\xspace}
\newcommand{\Sec}{{\ensuremath{\mathrm{s}}}\xspace}
\newcommand{\yr}{{\ensuremath{\mathrm{yr}}}\xspace}
\newcommand{\Msuns}{{\ensuremath{\Msun\,\Sec^{-1}}}\xspace}
\newcommand{\g}{{\ensuremath{\mathrm{g}}}\xspace}
\newcommand{\cm}{{\ensuremath{\mathrm{cm}}}\xspace}
\newcommand{\gcc}{{\ensuremath{\g\,\cm^{-3}}}\xspace}

\newcommand{\powersep}{{\ensuremath{\times}}}
\newcommand{\isofont}[1]{{\mathrm{#1}}}
\newcommand{\isomass}[1]{{\ensuremath{\isofont{^{#1}}}}}
\newcommand{\isocharge}[1]{{\ensuremath{\isofont{_{#1}}}}}
\newcommand{\isotope}[3]{{\ensuremath{\isocharge{#1}\isomass{#2}\isofont{#3}}}}
\newcommand{\I}[2]{{\isotope{}{#1}{#2}}}
\newcommand{\Ep}[1]{{\ensuremath{10^{#1}}}}
\newcommand{\E}[1]{{\ensuremath{\powersep\Ep{#1}}}}

\title{Evolution and Nucleosynthesis of Very Massive Primordial Stars}

\author{A.~Heger\address{Department of Astronomy and Astrophysics, Uiversity of California, Santa Cruz, CA~95064, U.S.A.},
        I.~Baraffe\address{C.R.A.L (UMR 5574 CNRS), Ecole Normale Sup\'erieure, 69364 Lyon Cedex 07, France},
	C.~L.~Fryer$^\mathrm{a}$\address{T-6 Group, LANL, Los Alamos, NM 87454, U.S.A.}, and
	S.~E.~Woosley$^\mathrm{a}$}
       
\begin{document}

% typeset front matter
\maketitle

\begin{abstract}
We investigate the evolution, final fate, and nucleosynthetic yields
of rotating and non-rotating very massive stars (VMS) of zero
metallicity.  First we address the issue of mass loss during hydrogen
burning due to vibrational instabilities.  We find that these objects
are much more stable than what was found in previous studies of VMS of
solar composition, and expect only negligible mass loss driven by the
pulsations.  As these stars thus reach the end of their evolution with
massive helium cores, they encounter the pair-creation instability.
We find that for helium core masses of $\sim64\ldots133\,\Msun$ these
stars are completely disrupted with explosion energies of up to
$\sim\Ep{53}\,\erg$ and eject up to $\sim60\,\Msun$ of \I{56}{Ni}.
Stars with more massive helium cores collapse into black holes.  We
present the first calculations that follow the collapse of such a
massive rotating star and predict that X-ray burst and significant
gravitational wave emission could result.
\end{abstract}

\section{INTRODUCTION}

Recently, three-dimensional cosmological simulations have reached
sufficient resolution on small scales to begin to address the star
formation problem of the first generation of so-called Population III
(Pop III) stars \cite{OG96,abe98}.  Though the formation and nature of
the first generation of metal-free stars have been investigated since
over thirty years \cite[and many others]{SS53,Yon72,HC77,PSS83,Sil83},
the extent to which they differed from present day stars in ways other
than composition is still debated.  While considerable uncertainty
remains, these simulations suggest that the first generation of stars
may have been quite massive, $\sim 100-1000\,\Msun$
\cite{Lar99,ABN00,BCL99}, giving us motivation to
examine the properties, evolution, and fate of such stars.

\section{VIBRATIONAL (IN-)STABILITY AND MASS LOSS}

Above a critical mass, main sequence stars are vibrationally unstable
due to the destabilizing effect of nuclear reactions in their central
regions ($\epsilon$-mechanism) \cite{Led41,SH59}.  According to
non-linear calculations, such an instability leads to mass loss rather
than catastrophic disruption \cite{App70,Tal71,Pap73}.  Previous
investigations of stars of solar-like compositions or slightly
metal-poor stars \cite{SH59,SS68,AHR75,Sto92,GK93,KFG93} indicated
that stars above a few 100\,\Msun would lose a substantial amount of
mass during hydrogen burning.  However, the structure of metal-free
stars is significantly different from those that even contain a trace
of initial metals.  We have thus performed, to our knowledge, the
first stability analysis of zero metallicity stars with $100\,\Msun
\le M \le 500\,\Msun$.

We find that the different structure and the higher central
temperatures of these stars make them less vibrationally unstable,
\textsl{i.e.}, the timescale for the growth of the amplitude is much
longer, and thus, based on the non-linear calculations of
\cite{App70}, we derive considerably lower mass loss rates than for
more metal rich counterparts, and find that even our 500\,\Msun star
loses only a negligible amount of mass during central hydrogen
burning.  During central helium burning only stars with initial masses
above $\sim300\,\Msun$ become pulsationally unstable during the last
few 10,000\,\yr of their evolution.  Therefore we estimate that stars
of $\lesssim500\,\Msun$ can even retain their hydrogen envelope until
they encounter the pair creation instability.  A more detailed
description of our analysis can be found in \cite{BHW00}.

\section{NUCLEOSYNTHETIC YIELDS OF PAIR CREATION SUPERNOVAE}

\begin{figure}[tb]
\includegraphics[width=\columnwidth]{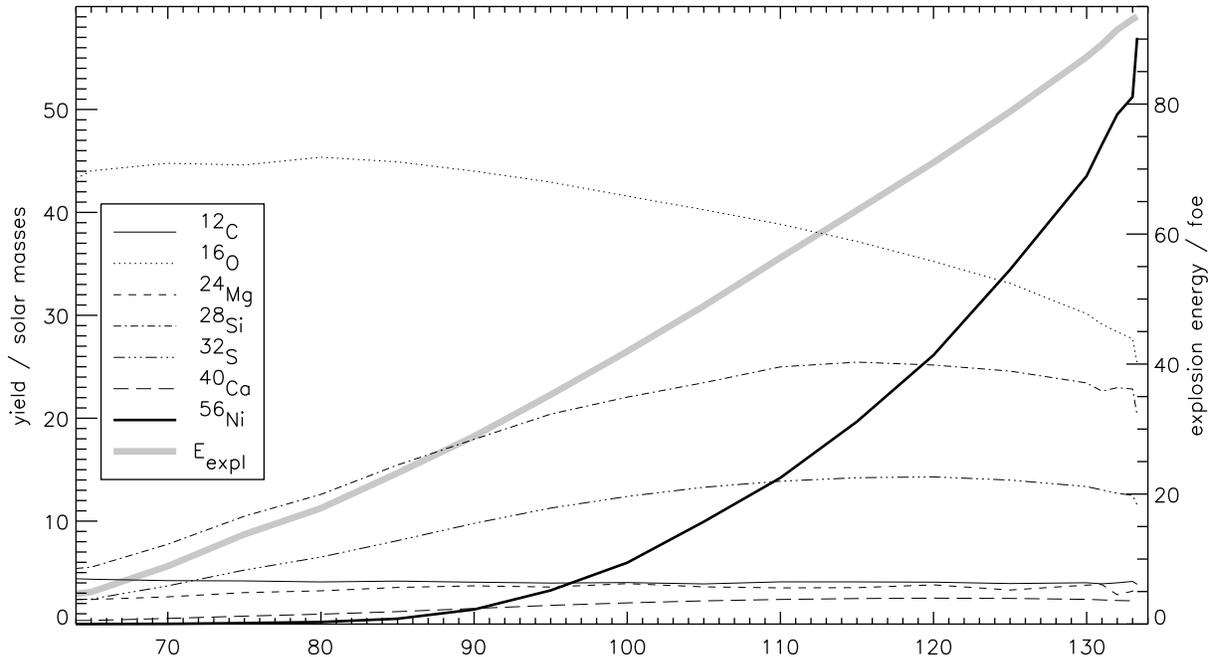}
\caption{%
Post-explosive nucleosynthetic yields of helium cores as a function of
mass (\textsl{black lines; left axis}).  \I{20}{Ne} and \I{36}{Ar}
have yields similar to those of respectively \I{24}{Mg} and
\I{40}{Ca}.  The \textsl{thick gray line} gives the explosion energy
in ``foe'' ($1\,\mathrm{foe}=\Ep{51}\,\erg$, about the explosion
energy of a ``typical'' core collapse supernova; \textsl{right axis}) .
\label{fig:yield}}
\end{figure}

The evolution of very massive stars has already been followed by many
authors before \cite[e.g.]{BAC84}.  We followed the evolution and
nucleosynthesis of rotating and non-rotating VMS Pop III stars from
onset of hydrogen burning through the electron-positron pair creation
instability until one year after the explosion \cite{HW00}, using
recent stellar ``input physics'' and extended nuclear reaction
networks.

In Fig.~\ref{fig:yield} we give the results of calculations of
non-rotating plain helium cores evolved without mass loss.  It
displays the yields of the ejecta as a function of helium core mass
and the resulting explosion energy.  For a
\I{12}C($\alpha$,$\gamma$)\I{16}O rate of $1.7\times$ that of
\cite{CF88}, in the extreme case of a $133.3\,\Msun$ helium core we
find an explosion energy of about 9.4\E{52}\,\erg and ejection of
57\,\Msun of \I{56}{Ni}, almost half of all ejecta.  The ejecta from
these objects are dominated by ``$\alpha$-nuclei'', but some nickel
isotopes are made in the $\alpha$-rich freeze-out in the center of the
most massive cores.  Essentially no elements above the iron group are
produced.  Below the iron group, the elements of even charge number
are produced in about solar abundance ratios, while elements of odd
mass number are strongly underproduced.

Below $\sim64\,\Msun$ we do not find prompt explosions of the helium
cores, but after a strong pair-instability induced pulse the outer
layers are ejected while the central parts of the star fall back and
contract again and are either disrupted in subsequent pulses or evolve
through to iron core collapse \cite{Woo86} and could become a
collapsar, similar to the models of \cite{MWH00}.  Above
$\sim133\,\Msun$ the photo-disintegration of heavy elements into
$\alpha$-particles, and $\alpha$-particles into nucleons, consumes so
much energy that the center of the star continues to collapse into a
black hole (see below).  Only in the mass range shown in
Fig.~\ref{fig:yield}, which corresponds to initial stellar mass of
$\sim150\ldots250\,\Msun$, the star is completely disrupted in one
pulse.

\section{SIMULATIONS OF THE COLLAPSE TO A BLACK HOLE}

We have followed the collapse of a rotating 300\,\Msun stars that had
a rotation of 10\,\% Keplerian on the zero-age main sequence and which
formed a $\sim180\,\Msun$ helium core.  The 1D model was mapped into a
2D Lagrangian SPH code when the central density exceeded 5\E{10}\,\gcc
and the further evolution was followed including, for the first time,
neutrino trapping during the collapse.  Sufficient angular momentum
was present for triaxial deformations to grow significantly in the
``proto-black hole'', with the possible consequence of a strong
gravity wave signal.  After the formation of the initial black hole,
it accreted mass to $\sim140\,\Msun$ at a rate of
$10\ldots100\,\Msuns$.  At this point, a centrifugally supported disk
forms that may produce jets along the poles through
magneto-hydrodynamical effects \cite{BZ77}.  Though the efficiencies of
the mechanisms under consideration are still very speculative, such a
jet could lead either to an energetic, jet-driven explosion or an
x-ray transient, depending on whether the hydrogen-rich envelope of
the star is still present or has been lost to a possible binary
companion star.  A more detailed description is given in \cite{FWH00}.

\section{SUMMARY AND CONCLUSIONS}

Recent results on the formation of Pop~III stars indicate that very
massive stars of zero metallicity could have formed.  Our analysis of
the pulsation properties indicates that stars of $\lesssim500\,\Msun$
are far less vibrationally unstable than previously though and thus
may reach the end of their evolution as objects with massive helium
cores that encounter the pair-creation instability.  This has
interesting consequences for the nucleosynthetic yield from these
objects, and may even allow for the observation of their explosions
despite the high red shift.  Above an initial mass of $\sim300\,\Msun$
the core collapses into a black hole and may cause an X-ray burst or a
strong gravity wave signal that may be accessible to experiments in
the near future.

The survival of these objects as very massive stars until the end of
their evolution also has interesting consequences for the
re-ionization of the early universe, as these VMS have an effective
surface temperature of $\sim\Ep5\,\K$ \cite{EC71,BHW00}, which means
that they very efficiently turn nuclear energy into ionizing photons.

\textbf{Acknowledgements.\/}
This research was supported, in part, by Prime Contract
No. W-7405-ENG-48 between The Regents of the University of California
and the United States Department of Energy, the National Science
Foundation (AST 97-31569, INT-9726315), and the Alexander von
Humboldt-Stiftung (FLF-1065004).

\end{document}